\documentclass[prd,twocolumn,showpacs,nofootinbib,preprintnumbers,floatfix]{revtex4}  
\usepackage[plainpages=false, colorlinks=true, anchorcolor=blue, linkcolor=blue, citecolor=blue, bookmarks=false]{hyperref}
\usepackage{graphicx}  
\usepackage{dcolumn}   
\usepackage{bm}        
\usepackage{amssymb}   
\usepackage{natbib}
\usepackage{amsmath}   
\usepackage{longtable}  
\usepackage{float} \usepackage[utf8]{inputenc}
\usepackage{tasks}

\begin{document}

\newcommand{\apjl}{Astrophys. J. Lett.}
\newcommand{\apjs}{Astrophys. J. Suppl. Ser.}
\newcommand{\aap}{Astron. \& Astrophys.}
\newcommand{\rthis}[1]{\textcolor{black}{#1}}
\newcommand{\aj}{Astron. J.}
\newcommand{\pasp}{PASP}
\newcommand{\araa}{Ann. Rev. Astron. Astrophys. } 
\newcommand{\aapr}{Astronomy and Astrophysics Review}
\newcommand{\ssr}{Space Science Reviews}
\newcommand{\mnras}{Mon. Not. R. Astron. Soc.}
\newcommand{\apss} {Astrophys. and Space Science}
\newcommand{\jcap}{JCAP}
\newcommand{\na}{New Astronomy}
\newcommand{\pasj}{PASJ}
\newcommand{\pasa}{Pub. Astro. Soc. Aust.}
\newcommand{\physrep}{Physics Reports}

\title{Probing the dark matter density evolution law with large scale structures}

\author{Kamal Bora$^{1}$}\email{ph18resch11003@iith.ac.in}

\author{R. F. L. Holanda$^{2,3,4}$}\email{holandarfl@fisica.ufrn.br}

\author{Shantanu Desai$^{1}$}\email{shntn05@gmail.com}

\affiliation{$^1$ Department of Physics, Indian Institute of Technology, Hyderabad, Kandi, Telangana-502285, India }

\affiliation{$^2$Departamento de F\'{\i}sica, Universidade Federal do Rio Grande do Norte,Natal - Rio Grande do Norte, 59072-970, Brasil}

\affiliation{$^3$Departamento de F\'{\i}sica, Universidade Federal de Campina Grande, 58429-900, Campina Grande - PB, Brasil}

\affiliation{$^4$Departamento de F\'{\i}sica, Universidade Federal de Sergipe, 49100-000, Aracaju - SE, Brazil}

\begin{abstract}
We propose a new method to explore a possible departure from the  standard time evolution law for the dark matter density. We looked for a violation of this law  by using a  deformed evolution law, given by $\rho_c(z) \propto (1+z)^{3+\epsilon}$, and then  constrain $\epsilon$. The dataset  used for this purpose consists of Strong Gravitational Lensing data obtained from SLOAN Lens ACS, BOSS Emission-line Lens Survey, Strong Legacy Survey SL2S, and SLACS; along with galaxy cluster  X-ray gas mass fraction  measurements obtained using the Chandra Telescope. Our analyses show that  $\epsilon$ is consistent with zero within 1 $\sigma$ c.l., but the current dataset cannot  rule out with high confidence level interacting models of dark matter and dark energy.
\end{abstract}
\pacs{98.80.-k, 95.35.+d, 98.80.Es}

\maketitle

\section{Introduction}
The current concordance model of the universe, consisting of about 25\% non-interacting cold dark matter, 70\% Dark energy and 5\% ordinary baryons agrees very well with Planck CMB observations~\cite{Planck18}. However, this concordance model still has a number of lingering issues,  such as the core-cusp and missing satellite problems at small scales, failure to detect Cold Dark Matter in the laboratory, Lithium-7 problem in Big-Bang Nucleosynthesis, Cosmological Constant problem, Cosmic coincidence problem, Hubble constant and $\sigma_8$ tension, etc (see for eg.~\cite{Caldwell2009,Weinberg2013,Bullock,Merritt,Fields,Divalentino21} for reviews of these problems). Therefore, a large number of alternative models to the standard $\Lambda$CDM cosmology have been studied to address some of these issues~\cite{alternatives}.

One possible solution to tackle the cosmic coincidence problem~\cite{Griest} is to introduce an interaction between the dark sectors of the universe~\cite{Wang2016}.
Several studies have previously been done in the past to explore the interaction between the dark energy and dark matter by positing an energy exchange between them~\cite{Ozer1985,amendola2000,barrow06,caldera09,alcaniz12}. In addition to this,~\citet{alcaniz12} also searched for these interactions using the current observations of the Large Scale Structure (LSS), Cosmic Microwave Background anisotropies, BAO measurements, and SNe Ia Hubble diagram. Recently, a plethora of studies have been undertaken in order to understand the dark sector interactions in a model-dependent as well as model-independent fashion, and some of these studies  hint towards deviations from the $\Lambda$CDM model at low redshifts, that might be associated with the Hubble tension~\cite{marttens19,marttens20a,marttens20b,carneiro19,clark21,supriya20,yang19,valentino17,vattis19,bonilla21,Holanda:2019sod}.

In this letter, we propose a new method to study a possible deviation from the  standard evolution law for the  dark matter density ($\rho_c(z) \propto (1+z)^3$) using the Strong Gravitational Lensing (SGL) data obtained from SLOAN Lens ACS+BOSS Emission-line Lens Survey (BELLS)+Strong Legacy Survey SL2S+SLACS along with X-ray gas mass fraction data from~\citet{Mantz:2014xba}.
In order to study any departure from the  standard evolution law, an ad-hoc term ($\epsilon$) is added to the cubic exponent, which is a function of the cosmic scale factor i.e. $\epsilon(a)$, that arises due to the non-gravitational interaction between the dark sectors. The modified evolution law of dark matter can therefore be written as, $\rho_c(z) \propto (1+z)^{3+\epsilon}$~\cite{wang05,alcaniz05}.

This paper is organized as follows. Section~\ref{methodology} explains the methodology adopted in this work. In \textbf{Section~\ref{data}}, we present the data sample used for our analysis. Section~\ref{sec:analysis} describes our analysis and results. Our conclusions are presented in Section~\ref{sec:conclusions}. 

\section{Methodology}
\label{methodology}

In this section, we discuss some aspects of SGL systems and gas mass fractions, and discuss how is it possible to combine these observations in order to put constraints on possible departures from standard evolution law for the dark matter density. 

\subsection{Strong Gravitational Lensing Systems}

SGL systems are one of the cornerstone predictions of  general relativity~\cite{lentes}.  Strong lensing is a purely gravitational phenomenon and  can be used to investigate gravitational and cosmological theories as well as  fundamental physics. Usually,  a lens could  be a foreground galaxy or  a cluster of galaxies positioned between a source (quasar) and an observer, where the multiple-image separation from the source only depends on the lens and source angular diameter distance (see, for instance,  Refs. \cite{Cao2015,Holanda:2016msr,Rana2017,Rana2017b,Ruan2018,Cao2018,Liao2019,Leaf2018,Amante:2019xao,Lizardo:2020wxw}, where SGL systems were used recently as a cosmological tool). However, it is important to point out that  the constraints obtained from SGL systems  may depend on a model for the lens mass distribution (see next section). With the simplest model assumption, the so called singular isothermal sphere (SIS) model,  one defines the Einstein radius ($\theta_E$), which is given by \cite{Bartelmann2010,Cao2015}:
\begin{figure}[t]
    \centering
    \includegraphics[width=10cm, height=8cm]{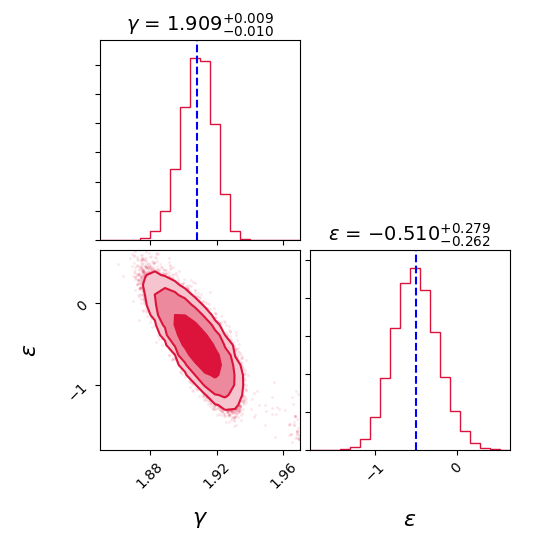} 
    \caption{\textbf{For Low mass range sample:} The 1D marginalized likelihood distributions along with 2D marginalized constraints showing the 68\%, 95\%, and 99\% credible regions for the parameters $\gamma$ and $\epsilon$, obtained using the {\tt Corner} python module~\cite{corner}.}
   \label{fig:low}
    
\end{figure}

\begin{figure}[t]
    \centering
    \includegraphics[width=10cm, height=8cm]{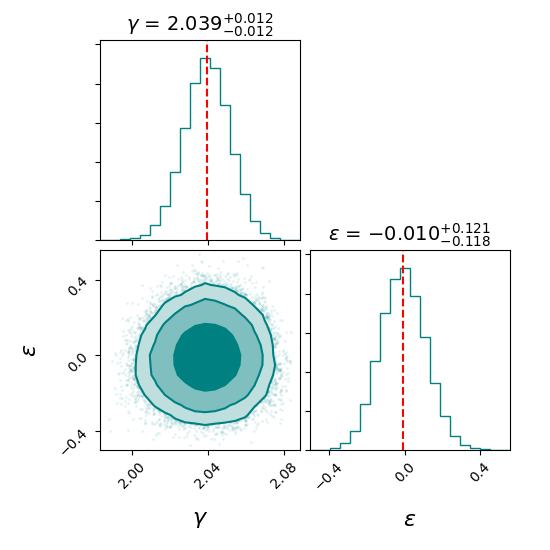} 
    \caption{\textbf{For Intermediate mass range sample:} The 1D marginalized likelihood distributions along with 2D marginalized constraints showing the 68\%, 95\%, and 99\% credible regions for the parameters $\gamma$ and $\epsilon$, obtained using the {\tt Corner} python module~\cite{corner}.}
    \label{fig:intermediate}
\end{figure}

\begin{figure}[t]
    \centering
    \includegraphics[width=10cm, height=8cm]{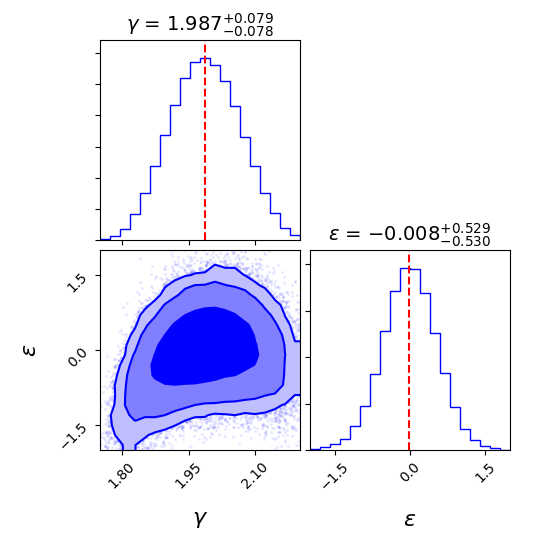} 
    \caption{\textbf{For High mass range sample:} The 1D marginalized likelihood distributions along with 2D marginalized constraints showing the 68\%, 95\%, and 99\% credible regions for the parameters $\gamma$ and $\epsilon$, obtained using the {\tt Corner} python module~\cite{corner}.}
    \label{fig:high}
\end{figure}

\begin{equation}
\theta_E = 4\pi \frac{D_{A_{ls}}}{D_{A_{s}}} \frac{\sigma_{SIS}^{2}}{c^2}
\label{eq:thetaE_SIS}
\end{equation}
In this equation,  $D_{A_{ls}}$ is the angular diameter distance from the lens to the source, $D_{A_{s}}$ the angular diameter distance of the observer to the source, $c$ the speed of light, and $\sigma_{SIS}$ the velocity dispersion caused by the lens mass distribution. 

In our method, we assume a flat universe and use the following  observational quantity from the SGL systems \cite{Liao2019}: 
\begin{equation}
\label{eq:D_SIS}
D={\frac{D_{A_{ls}}} {D_{A_s}}}=\frac{{\theta}_E c^2}{4{\pi} \sigma^2_{SIS}}
\end{equation}
In a flat universe, the comoving distance $r_{ls}$ is given  by \cite{Bartelmann2010}
$r_{ls}=r_s-r_l$, and using $r_s=(1+z_s)D_{A_s}$, $r_l=(1+z_l)D_{A_l}$  and $r_{ls}=(1+z_s)D_{A_{ls}}$, we find
\begin{equation}
\label{eq:D}
D= 1 - \frac{(1+z_l)D_{A_{l}}}{(1+z_s)D_{A_{s}}}
\end{equation}
Finally, by using the cosmic distance duality relation
 $D_L=(1+z)^2D_A$~\cite{CDDR,boraCDDR21}, Eq.~\ref{eq:D} can be written as
\begin{equation}
\label{eq:final_D}
\frac{(1+z_s)}{(1+z_l)}= (1-D)\frac{D_{L_s}}{D_{L_l}}
\end{equation}

\subsection{Gas mass fraction}

The cosmic gas mass fraction can be  defined as $f_{gas} \equiv \Omega_b/\Omega_M$ (where $\Omega_b$ and $\Omega_M$ are the baryonic and total matter density parameters, respectively), and the constancy of this quantity within massive, relaxed clusters at $r_{2500}$ can be used to constrain cosmological parameters by using the following equation (see, for instance, \cite{Allen2007,Ettori2009,Allen2011,Mantz:2014xba,Holanda2020}) 
\begin{equation}
\label{fgas}
f_{gas}(z) = N \left[\frac{\Omega_b(z)}{\Omega_b(z) +\Omega_{c}(z)}\right] \left(\frac{D_L^*}{D_L}\right)^{3/2}\
\end{equation}
Here,  the observations are done in the X-ray band, the asterisk denotes the corresponding quantities for the fiducial model used in the observations to obtain  $f_{gas}$ (usually a flat $\Lambda$CDM model with Hubble constant $H_0=70$ km s$^{-1}$ Mpc$^{-1}$ and the present-day total matter density parameter $\Omega_M=0.3$), $\Omega_{c}(z)$ is the dark matter density parameter, the normalization factor $N$ carries all the astrophysical information about the matter content in the cluster, such as stellar mass fraction,  non-thermal pressure and the depletion parameter $\gamma$, which indicates the amount of cosmic baryons that are thermalized within the cluster potential (see details in the Refs. \cite{Allen2007,Mantz:2014xba,Battaglia2013,Planelles2013}). The ratio in the parenthesis of Eq.~\ref{fgas} encapsulates the expected variation in $f_{gas}$ when the underlying cosmology is varied, which makes the analyses with gas mass fraction measurements model-independent. Finally, it is important to stress that the Eq.~\ref{fgas} is obtained only when the cosmic distance duality relation is  valid (see Ref.\cite{Gon2012} for details).

The key equation to our method can be  obtained when one combines Equations \ref{eq:final_D} and \ref{fgas} by taking into account a possible departure from the dark matter density standard evolution law, such as $\Omega_c(z)=\Omega_{c0}(1+z)^{3+\epsilon}$. In this way, we now obtain:

\begin{equation}
\label{main_equation}
\left[\frac{\rho_{b0}+\rho_{c0}(1+z_l)^{\epsilon}}{\rho_{b0}+\rho_{c0}(1+z_s)^{\epsilon}}\right] =  \left[\frac{(1+z_s)D^*_{L_l}}{(1+z_l)D^*_{L_s}}\right]^{3/2}\left[\frac{f_{gas}(z_s)}{f_{gas}(z_l)}\right](1-D)^{-3/2}
\end{equation}
 As one may see, unlike Ref.\cite{Holanda:2019sod}, where  the gas mass fraction measurements and SNe Ia luminosity distance were also used to obtain limits on $\epsilon$, our results are independent from the baryon budget for the  clusters, as long as  the $N$ factor does not depend  upon the  redshift of the cluster \cite{2013ApJ...777..123B,2013MNRAS.431.1487P,Holanda:2017cmc,Bora:2021bui}.


\section{Cosmological data}
\label{data}
We now describe in detail the data used for our analysis.
\begin{itemize}
\item We use the most recent X-ray gas mass fraction measurements of $40$ galaxy clusters in the redshift range $0.078 \leq z \leq 1.063$ from Ref. \cite{Mantz:2014xba}.  The data set employed here consists of Chandra observations, identified through a comprehensive search of the Chandra archive for hot ($kT \geq  5$ keV), massive and morphologically relaxed systems. The restriction to relaxed systems minimizes the systematic biases
due to departures from hydrostatic equilibrium and substructure, as well as the scatter due to these effects, asphericity, and projection. The aforementioned work incorporated a robust gravitational lensing calibration of the X-ray mass estimates \cite{app} and restricted the  measurements to the most self-similar and accurately measured regions of clusters. Therefore the  systematic uncertainties were significantly reduced when compared to previous works in literature. The gas mass fractions were obtained from spherical shells at radii near $r_{2500}$, rather than the cumulative fraction integrated over all radii ($< r_{2500}$). As stressed in Ref.~\cite{Mantz:2014xba}, a consequence of the use of spherical shells (excluding the cluster core) is that
it is possible to directly use the   simulated results for the gas depletion, rather than combining a prior on the baryonic depletion with measurements of the mass in stars relative to the hot gas, without incurring an additional systematic uncertainty.
 From the lowest-redshift data  in their sample (consisting of  five clusters at $z <  0.16$), they obtained a constraint on a combination of the Hubble parameter and the cosmic baryon fraction, such as: $\frac{h^{3/2}\Omega_{b0}}{\Omega_{C0}+\Omega_{b0}} = 0.089 \pm 0.012$, insensitive to the nature of dark energy \cite{Mantz:2014xba}. Then, by combining this with the values of $h$ ($h= 0.732 \pm 0.013$) \cite{2021arXiv210301183D,Riess:2020fzl} and $100 \Omega_{b0}h^2 \text{(from BBN)} = 2.235 \pm 0.033$ \cite{Cook2018}, we obtain: $\rho_{b0}=4.20 \pm 0.22 (\times 10^{-31} gm/cm^{3})$ and $\rho_{c0}=25.34 \pm 4.35(\times10^{-31}  gm/cm^3)$. These values will be used in our analyses as  approximate local estimates.
 
\item We also consider subsamples  from a specific catalog containing 158 confirmed sources of strong gravitational lensing \cite{Leaf2018}. This complete compilation includes 118 SGL systems identical to the compilation of \cite{Cao2015}, which was obtained from a combination of SLOAN Lens ACS, BOSS Emission-line Lens Survey (BELLS), and Strong Legacy Survey SL2S, along with 40 new systems recently discovered by SLACS and pre-selected by \cite{2017ApJ...851...48S} (see Table I in \cite{Leaf2018}).  For the mass distribution of lensing systems, the so-called power-law model is considered. This one assumes a spherically symmetric mass distribution with a more general power-law index $\gamma $, namely $\rho \propto r^{-\gamma}$ (several studies have shown that the slopes of density profiles of individual galaxies show a non-negligible deviation from the SIS \cite{Koopmans2009,Auger2010,Barnab2011,Sonnenfeld2013,Cao2016,HolandaSaulo,Chen2019}). In this approach $\theta_E$ is given by:
\begin{equation}
\theta_E =   4 \pi
\frac{\sigma_{ap}^2}{c^2} \frac{D_{ls}}{D_s} \left[
\frac{\theta_E}{\theta_{ap}} \right]^{2-\gamma} f(\gamma),
\label{thetaE}
\end{equation}
where $\sigma_{ap}$ is the  stellar velocity dispersion inside an aperture of size $\theta_{ap}$ and
\begin{eqnarray} \label{f factor}
f(\gamma) &=& - \frac{1}{\sqrt{\pi}} \frac{(5-2 \gamma)(1-\gamma)}{(3-\gamma)} \frac{\Gamma(\gamma - 1)}{\Gamma(\gamma - 3/2)}\nonumber\\
          &\times & \left[ \frac{\Gamma(\gamma/2 - 1/2)}{\Gamma(\gamma / 2)} \right]^2
\end{eqnarray}
Thus, we obtain: 
\begin{equation} 
\label{NewObservable}
 D=\frac{D_{A_{ls}}}{D_{A_{s}}} = \frac{c^2 \theta_E }{4 \pi \sigma_{ap}^2} \left[ \frac{\theta_{ap}}{\theta_E} \right]^{2-\gamma} f^{-1}(\gamma)
\end{equation}
For $\gamma = 2$, we recover the singular isothermal spherical distribution. The relevant information necessary to obtain $D$ can be found in Table 1 of \cite{Leaf2018}.  The complete data (158 points) is reduced to  98 points, whose redshifts are lower than $z = 1.061$ and with the quantity $D \pm \sigma_D$ (by taking $\gamma=2$)  lower than the unity  ($D > 1$ represents a non physical region).  Our compilation contains only those systems with early type galaxies acting as lenses, with spectroscopically measured stellar apparent velocity dispersion, estimated apparent and Einstein radius, and both the lens and source redshifts.

However, the cosmological analyses by using SGL systems are strongly dependent on the density profile describing the mass distribution of gravitational lensing systems. Recent papers have explored a possible redshift evolution of the mass density power-law index \cite{Cao2016,HolandaSaulo,Amante:2019xao,Chen2019}. No significant evolution has been found. However, the results suggest that it is prudent to  treat low, intermediate and high-mass galaxies separately in analyses. As commented by  Ref.\cite{Cao2016}, elliptical galaxies with velocity dispersions smaller than $200$ km/s may be classified roughly as relatively low-mass galaxies, while those with velocity dispersion larger than $300$ km/s may be treated as relatively high-mass galaxies. Naturally, elliptical galaxies with velocity dispersion between $200-300$ km/s may be classified as intermediate-mass galaxies. In this way, in our analyses we work with three  sub-samples consisting of: 26, 63, and 9 data points with low, intermediate, and high $\sigma_{ap}$, respectively.

As one may see, in order to put limits on $\epsilon$ by using the Eq.~\ref{main_equation}, it is necessary to have  gas mass fraction measurements  at the  lens and source redshifts, for each SGL system. These quantities are calculated by applying Gaussian Process using the 35 gas mass fraction measurements compiled by  Ref.\cite{Mantz:2014xba} (here, the 5 clusters with $z<0.16$ were excluded).
\end{itemize}
\vspace{0.2cm}

\begin{figure*}[t]
    \centering
    \includegraphics[scale=0.33]{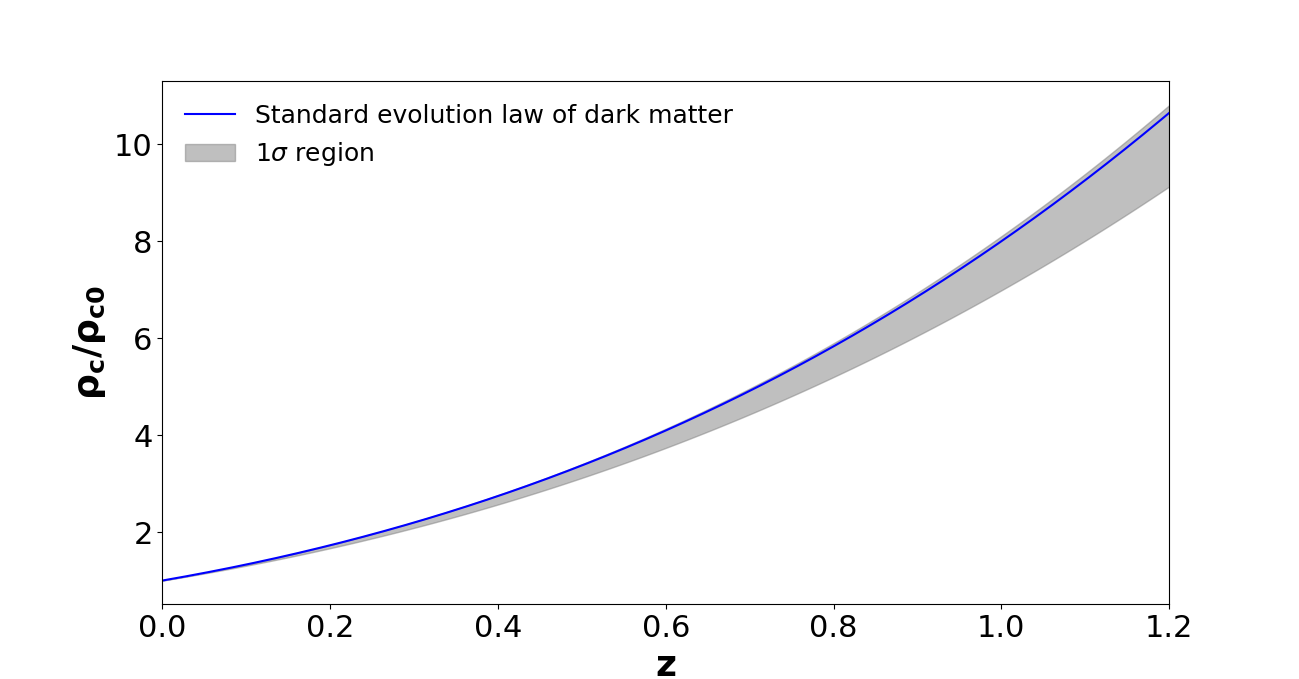} 
    \caption{The evolution law for the density of  dark matter as a function of redshift. The dark matter density is normalized by  $\rho_{c0}$, which  represents the dark matter density at $z = 0$. The grey shaded area shows the  $1\sigma$  allowed region for  the evolution law as found in this work. The standard evolution law for dark matter is displayed by the blue line.  }
    \label{fig:f4}
\end{figure*}

\section{Analysis and Results} 
\label{sec:analysis}

The constraints on the $\gamma$ and $\epsilon$ parameters can be obtained by maximizing the likelihood distribution function, ${\cal{L}}$  given by

\begin{widetext}
\begin{equation}
    \label{eq:logL}
   -2\ln\mathcal{L} = \sum_{i=1}^{n} \ln 2\pi{\sigma_{i}^2}+ \sum_{i=1}^{n}\frac{\left(\zeta(\epsilon,z_i)- \left[\frac{(1+z_s)D^*_{L_l}}{(1+z_l)D^*_{L_s}}\right]^{3/2}\left[\frac{f_{gas}(z_s)}{f_{gas}(z_l)}\right](1-D)^{-3/2}\right)^2}{\sigma_{i}^2} , 
\end{equation}          \end{widetext}

where 
\begin{equation}
  \zeta(\epsilon,z_i) =  \left[\frac{\rho_{b0}+\rho_{c0}(1+z_l)^{\epsilon}}{\rho_{b0}+\rho_{c0}(1+z_s)^{\epsilon}}\right]
\end{equation}

\begin{table*}[]
\caption{\label{tab:table1}. Constraints on the parameters $\gamma$ and $\epsilon$ for different $\sigma_{ap}$ range used in this analysis as discussed in Sect.~\ref{sec:analysis} .}
    \centering
    \begin{tabular}{|l|c|c|c|r|} \hline
    \textbf{Sample} & \boldmath$\sigma_{ap}$\textbf{(km/sec)} & \boldmath$\gamma$ & \boldmath$\epsilon$\\ \hline 
      Low\ & $\sigma_{ap}< 200$ & $1.909^{+0.009}_{-0.010}$  &  $-0.510^{+0.279}_{-0.262}$ \\
      Intermediate & $200 < \sigma_{ap}< 300$ & $2.039\pm0.012$ & $-0.010^{+0.121}_{-0.118}$ \\
      High &  $\sigma_{ap}> 300$  & $1.987^{+0.079}_{-0.078}$ & $-0.008^{+0.529}_{-0.530}$\\
      Intermediate+High &  $200 < \sigma_{ap} \leqslant 396$  & $2.038\pm0.012$ & $-0.003^{+0.117}_{-0.115}$\\
      
      \hline 
      
    \end{tabular}

\end{table*}
Here, $\sigma_i$  denotes the statistical errors associated with the gravitational lensing  observations and gas mass fraction measurements, and are obtained by using standard propagation errors techniques. In the very first analyses using the method proposed here, let us fix  the $\rho_{b0}$ and $\rho_{c0}$ quantities to their best fit estimates discussed in the previous section: $4.20 \pm 0.22 (\times10^{-31}  gm/cm^3$) and  $25.34 \pm 4.35(\times10^{-31}  gm/cm^3$), respectively.

Now, we maximize our likelihood function with the help of $\tt{emcee}$ MCMC sampler~\cite{emcee} in order to estimate the free parameters used in Eq.~\ref{eq:logL}, viz. $\gamma$ and $\epsilon$. The one-dimensional marginalized posteriors for each parameter along with the 68\%, 95\%, and 99\% 2-D marginalized credible intervals, are shown in Fig.~\ref{fig:low}, Fig.~\ref{fig:intermediate}, and Fig.~\ref{fig:high} for the Low, Intermediate, and High samples, respectively. As we can see, the low and intermediate mass SGL sub-samples are not compatible with the SIS model ($\gamma=2$) even at 3$\sigma$ c.l. The high SGL sub-sample is in full agreement with the SIS model (see Table~\ref{tab:table1}). 
Moreover, the low sub-sample shows a non-negligible departure from standard evolution law while the intermediate and high sub-samples are in full agreement with the standard value ($\epsilon=0$).  We also perform a joint analysis by combining the intermediate and high mass samples. For this case, it is found that $\epsilon \approx 10^{-3}$, albeit with  large error bars. It is worth noting  that recent  cosmological estimates by using SGL systems with $\sigma_{ap} <  210$ km/sec were found to be in disagreement with SNe Ia  and CMB estimates (see Figures 1, 2, and 3 of Ref.~\cite{Amante:2019xao}). Then, if the $\epsilon$ result from this subsample is further confirmed by future and better SGL surveys it would bring to light a possible  evidence for new
Physics. We also  find that the degeneracy directions of $\epsilon$ and $\gamma$ change for each subsample. As one may see, our results also reinforce the need for  segregating the  lenses with low, intermediate and high velocity dispersions, and analyzing them separately in cosmological estimates. The results of all these analyses are summarized  in Table~\ref{tab:table1}. Since the likelihoods for the $\epsilon$ parameter from different subsamples are compatible with each other to  within about  1$\sigma$, we calculate an error-weighted average and found $\epsilon=-0.088 \pm 0.11$. In Fig.~\ref{fig:f4}, we show our observed trend of the dark matter density (and thereby $\epsilon$)  evolution by using the error-weighted average of our measurements, along with the standard expected evolution. As we can see,  the  standard evolution law (blue line)   is in full agreement with the 1$\sigma$ c.l. region found in this work. 

We now compare our results with those obtained in Ref~.\cite{Holanda:2019sod}. This work  discussed a model-independent way to obtain limits on the $\epsilon$  parameter  by combining the gas mass fraction measurements in galaxy clusters and  Type Ia
supernovae observations, and obtained $\epsilon=0.13 \pm 0.235$ ($1\sigma$ c.l.). However, their result depends on the $N$ factor (see Eq.~\ref{fgas}), while the results presented here are independent of the baryon budget of galaxy clusters (as long as $N$ is a constant).

\section{Conclusions}
\label{sec:conclusions}

In this letter we have proposed and carried out a  test, to probe the dark matter density time evolution law: $\rho_c(z) \propto (1+z)^{3+\epsilon}$ ($\epsilon =0$ recovers the standard law). Strong gravitational lensing systems (SGL) and gas mass fractions of galaxy clusters were used as the data set for this analysis.  The basic premises used in our analyses  were: the flat universe assumption and the validity of  cosmic distance duality relation. The lens profiles in SGL systems were described by a power law model ($\rho \propto r^{-\gamma}$), but the $\gamma$ parameter was not considered to be  universal for all the lens mass intervals. 
 
 By considering, separately, three sub-samples of SGL systems,  which differ from each other by their stellar velocity dispersion values, the combined analyses with gas mass fraction data showed a non-negligible departure of the standard law ($\epsilon \neq 0$) for the low mass sub-sample (see Table~\ref{tab:table1} and Fig.~\ref{fig:low}), while the intermediate and high sub-samples together indicate $\epsilon \approx 10^{-3}$ (see Table~\ref{tab:table1}). The likelihoods obtained for the $\epsilon$ parameter from different sub-samples are compatible with each other to  within about  1$\sigma$, then  an error-weighted average was calculated and we found $\epsilon=-0.088 \pm 0.11$. On the other hand, the SIS model ($\gamma=2$)  was compatible only with the high mass sub-sample. Our results do not  depend on the  baryon budget of galaxy clusters ($N$ factor in Eq.~\ref{fgas}). 
 
Therefore, we conclude that, although $\epsilon$ was found to be consistent with zero within 1 $\sigma$ c.l.,  the current constraints obtained here are unable to  confirm or rule out an interaction in the dark sector due the large error bars.

{ In principle,  constraints on $\epsilon$ parameter can also be obtained  using distance measurements from current and upcoming Baryon Acoustic Oscillation(BAO) measurements~\cite{ahumada20}. We plan to explore this in a future work. }
However, a more definitive test  for  a possible  non-gravitational interaction in the dark sector using the same method discussed here, should be possible from the X-ray survey eROSITA \cite{eRosita}, that is expected to detect $\approx$ 100,000 galaxy clusters, along with followup optical and infrared data from EUCLID mission, Vera Rubin LSST, and \textbf{Nancy Grace Roman} space telescope, that will discover thousands of strong lensing systems.

\section*{ACKNOWLEDGEMENT}
We are grateful to the anonymous referee for useful constructive feedback on the manuscript. KB would like to thank the Department of Science and Technology, Government of India for providing the financial support under DST-INSPIRE Fellowship program. RFLH
thanks CNPq No.428755/2018-6 and 305930/2017-6.

    

    

\bibliography{ref}

\end{document}